\documentclass[aps,prd,twocolumn,longbibliography]{revtex4-1}

\usepackage{graphicx}
\usepackage{dcolumn}
\usepackage{bm}
\usepackage{graphicx}
\usepackage{dcolumn}
\usepackage{bm}
\usepackage{bigints}
\usepackage{color}
\usepackage{amsmath,amssymb,graphicx}
\usepackage{float}
\usepackage{braket}
\usepackage{textcomp}

\AtBeginDocument{%
    \newwrite\bibnotes
    \def\bibnotesext{Notes.bib}
    \immediate\openout\bibnotes=\jobname\bibnotesext
    \immediate\write\bibnotes{@CONTROL{REVTEX41Control}}
    \immediate\write\bibnotes{@CONTROL{%
    apsrev41Control,author="08",editor="1",pages="1",title="0",year="1"}}
     \if@filesw
     \immediate\write\@auxout{\string\citation{apsrev41Control}}%
    \fi
}%

\begin{document}

\preprint{APS/123-QED}

\title{Group velocity distribution and short-pulse dispersion in a disordered transverse Anderson localization optical waveguide}

\author{Arash Mafi}%
 \email{mafi@unm.edu}
\affiliation{Department of Physics \& Astronomy, University of New Mexico, Albuquerque, New Mexico 87131, USA
             \\Center for High Technology Materials, University of New Mexico, Albuquerque, New Mexico 87106, USA}%

\date{\today}

\begin{abstract}
We investigate the group velocity distribution of waveguide modes in the presence of disorder.
The results are based on extensive numerical simulations of disordered optical waveguides using statistical methods. 
We observe that the narrowest distribution of group velocities is obtained in the presence of a small amount of disorder; therefore, the modal dispersion of an 
optical pulse is minimized when there is only a slight disorder in the waveguide. The absence of disorder or the presence of a large amount of disorder can result in 
a large modal dispersion due to the broadening of the distribution of the group velocities. We devise a metric that can be applied to the mode group index 
probability-density-function and predict the optimal level of disorder that results in the lowest amount of modal dispersion for short pulse propagation. 
Our results are important for studying the propagation of optical pulses in the linear regime, e.g., for optical communications; and the nonlinear regime for 
high-power short-pulse propagation.
\end{abstract}
\maketitle
\section{Introduction}
Transverse Anderson localization (TAL) of light was first proposed by Abdullaev, et al.~\cite{transverse-Abdullaev} and De Raedt, et al.~\cite{transverse-DeRaedt}
in a dielectric medium with a transversely random and longitudinally uniform refractive index profile. They showed that 
an optical beam can propagate freely in the longitudinal direction while being 
trapped (Anderson-localized~\cite{Anderson1,Anderson1980,Abrahams-50-book,Lagendijk-Physics-Today-2009,Abrahams-Scaling-Theory,Stone,sheng2006introduction}) in the disordered transverse direction(s).
TAL of light has since been observed in various optical systems with one or two transversely 
disordered dimensions~\cite{Schwartz2007,Lahini-1D-AL-2008,Martin-1D-AL-2011,Mafi-Salman-OL-2012,Mafi-Salman-OPEX-2012,Mafi-Salman-OMEX-2012,SegevNaturePhotonicsReview,Mafi-AOP-2015,fratini2015anderson,yao2017beam,yilmaz2019transverse,PhysRevA.99.063807}. 
In particular, Karbasi, et al. reported the first observation of TAL in disordered optical 
fibers~\cite{Mafi-Salman-OL-2012,Mafi-Salman-OPEX-2012,Mafi-Salman-OMEX-2012}. The disordered optical fibers have since been used for 
high-quality image transport~\cite{Mafi-Salman-Nature-2014,Tuan:18,tong2018characterization,zhao2018image,zhao2018deep}, beam multiplexing~\cite{Mafi-Salman-Multiple-Beam-2013}, 
wave-front shaping and sharp focusing~\cite{Mafi-Marco-singlemode-2017,Mafi-Marco-Nature-light-focusing-2014,Mafi-Behnam-Optica-2018}, 
nonlocal nonlinearity~\cite{Mafi-Marco-APL-self-focusing-2014,Mafi-Marco-PRL-Migrating-NL-2014}, 
single-photon data packing~\cite{Mafi-Marco-information-2016}, optical diagnostics~\cite{liang2018surface}, and random lasers~\cite{Mafi-Behnam-Random-Laser-2017,joshi2019effect}.

TAL optical fiber (TALOF) is essentially a {\em highly} multimode optical fiber (MMF) with a {\em transversely} random refractive index profile. 
What sets a TALOF apart from a conventional MMF is that its guided modes are spatially localized due to the transverse disorder, while the guided modes in a 
conventional MMF typically cover all or a large portion of the guiding region~\cite{okamoto2005fundamentals,gloge1973multimode,Feit:78}. The modal characteristics of MMFs 
are generally responsible for their performance for the desired functionality~\cite{jolivet2009beam,vcivzmar2012exploiting,papadopoulos2013high,amitonova2016high,zhu2010coherent}; e.g., the 
mean localization radius of the modes in an imaging TALOF determines the average 
point spread function (PSF) across the tip of the fiber, where a stronger localization leads to a narrower PSF and a higher resolution image transport~\cite{Mafi-Salman-Nature-2014}.
Similarly, the standard deviation in the localization radius of the modes determines the uniformity of the image transport across the fiber.
Because the refractive index profile of a TALOF is random and the guided modes are numerous, the modal characteristics of a 
TALOF must be studied statistically~\cite{Mafi-AOP-2015,Mafi-Salman-Modal-JOSAB-2013}. 
This stochastic nature of TALOFs and the diversity of the physical attributes of the
localized modes is the key differentiating factor between the linear/nonlinear dynamics observed in TALOFs versus conventional MMFs. 

The modal area statistics of disordered quasi-one-dimensional (quasi-1D) and quasi-two-dimensional optical waveguides
were studied recently by Abaie, et al.~\cite{Mafi-Behnam-Scaling-PRB-2016,Mafi-Abaie-OL-2018} using the mode-area probability-density-function (PDF).
The mode-area PDF characterizes the relative distribution of the mode-areas of the guided modes in a disordered waveguide.
In particular, Abaie, et al. showed that the mode-area PDF converges to a terminal configuration as the transverse dimensions of the waveguide are increased. 
Therefore, it may not be necessary to study a real-sized disordered structure to obtain its full statistical localization properties and the 
PDF can be obtained for a considerably smaller structure. This observation is not only important from a fundamental standpoint, it has practical 
implications because it can reduce the often demanding computational cost that is required to study the statistical properties 
of Anderson localization in disordered waveguides. 
We emphasize that the mode-area PDF encompasses all the relevant statistical information on spatial localization of the guided modes and is a powerful tool
for studying the TAL. In this paper, we employ a similar statistical analysis based on the {\em probability-density-function}
to study the dispersive properties of TAL in disordered waveguides.

In the modal language, the dispersive properties of a waveguide are determined by the frequency ($\omega$) dependence of the propagation constant, 
$\beta(\omega)$, of the guided modes~\cite{agrawal2013nonlinear,agrawal2012fiber}. Determining the optical dispersive properties of TALOFs is critical to the understanding of their
linear and nonlinear characteristics, or in the continuous wave (CW) or pulsed laser operation, a few examples of which are as follows.
For each mode labeled with an index $i$, the full form of $\beta_i(\omega)$ over a broad frequency range is needed to determine
the phase-matching wavelengths for the intermodal (nonlinear) four-wave mixing (FWM) process~\cite{stolen1974phase,lin1981large,hill1981efficient,Pourbeyram:15,Nazemosadat:13}. 
In some cases, the Taylor expansion of $\beta(\omega)$ around a central frequency of $\omega_0$ and the corresponding 
local frequency derivatives, $\beta^{(n)}_i=\partial^n_\omega\beta_i|_{\omega_0}$, are sufficient to characterize the dispersive properties of an optical fiber~\cite{agrawal2013nonlinear,agrawal2012fiber}.
For example, consider $\beta^{(1)}_i(\omega)$, which determines the group velocity associated with the mode labeled with the index $i$:
in the linear regime, the difference between group velocities of different modes in a multimode 
fiber is responsible for the intermodal dispersion, which is generally the main limiting factor for the achievable transmission bandwidth (data-rate) 
in a multimode optical fiber communications system~\cite{agrawal2012fiber,gloge1973impulse,fan2005principal}. 
When an optical pulse is sent through a MMF, the intermodal dispersion (different values of $\beta^{(1)}_i$ for different modes) 
causes the pulse to break into multiple sub-pulses, each propagating with a different group velocity. 
Therefore, the distribution of group velocities determines the achievable data-rate. For example, a nanosecond-long optical 
pulse is hardly affected by the propagation through a 1\,km-long high-quality graded index fiber~\cite{Olshansky:76,koike1995high}; however, the same pulse is highly distorted by the modal
dispersion in a comparable step-index optical fiber. $\beta^{(1)}_i(\omega)$ also determines the pulsed nonlinear dynamics, including that 
of soliton propagation in MMF~\cite{WiseA5,WiseA13,Raghavan}.
For CW laser applications, the cavity response, including the free spectral range (FSR), is determined by the values of
$\beta_i(\omega)$ for the relevant modes, which are excited in the lasing process; their distribution can dictate the spectrum 
of the laser via the optical Vernier effect~\cite{Vasdekis:07,EsmaeilSelectivity}. The pulsed dynamics of such lasers, including the Q-switching and mode-locking, are similarly controlled
by the values of $\beta^{(1)}_i(\omega)$ for the relevant modes~\cite{Mafi-Behnam-Random-Laser-2017}.
\subsection{Studying a quasi-1D disordered slab waveguide}
In this manuscript, we focus on the statistics of the groups velocity (GV) of the guided modes and determine the GV-PDF of disordered waveguides.
Understanding the GV distribution underlies a large number of dispersive phenomena in guided wave systems~\cite{ho2011statistics}. Solving for all the guided modes for a 
given TALOF and obtaining proper statistical averages over many fiber samples is a formidable task even for large computer clusters.
For example, the V-number of the disordered polymer TALOF in Ref.~\cite{Mafi-Salman-OL-2012} with an air cladding is approximately 2,200 at 405~nm wavelength
resulting in more than 2.3 million guided modes. Recall that the V-number is given by
\begin{align}
V=\dfrac{\pi t}{\lambda}\sqrt{n^2_{\rm co}-n^2_{\rm cl}},
\label{eq:V-number}
\end{align}
where $\lambda$ is the
optical wavelength, $t$ is the core diameter of the fiber (or the core width for the case of a quasi-1D slab waveguide), and $n_{\rm co}$
($n_{\rm cl}$) is the effective refractive index of the core (cladding). The total number of the bound guided modes in a step-index
optical fiber is $\approx V^2/2$. As such, in order to lay the groundwork for understanding the statistical behavior of GV distribution in TALOFs,
we have decided to present a comprehensive characterization of a quasi-1D Anderson localized optical waveguide in this manuscript. 
This exercise is quite illuminating as it sheds light on the general statistical behavior of GV distribution and shows the extent of information that can 
be extracted from such distributions. The detailed analysis of the TALOF structure will be presented in a future publication.
\subsection{Wave equation for the guided modes}
Here, we have chosen to calculate the transverse electric (TE) modes of the disordered 
waveguide using the finite element method (FEM) presented in Refs.~\cite{Mafi-Behnam-Boundary-OC-2016,El-Dardiry-microwave-2012,Kartashov-NL-2012,Lenahan}. 
Similar observations can be drawn for transverse magnetic (TM) guided modes, but we limit our analysis to TE in this paper for simplicity. 
The appropriate partial differential equation that will be solved in this manuscript is 
the Helmholtz equation for electromagnetic wave propagation 
in a z-invariant dielectric waveguide
\begin{align}
\label{eq:helmholtz}
\nabla^2_{\rm T}A({\bf x}_{\rm T})+n^2({\bf x}_{\rm T})k_0^2A({\bf x}_{\rm T})=\beta^2A({\bf x}_{\rm T}),
\end{align}
where $A({\bf x}_{\rm T})$ is the transverse profile of the (TE) electric field 
$E({\bf x}_{\rm T},z,t)=A({\bf x}_{\rm T})\exp(i \beta z-i\omega t)$, 
which is assumed to propagate freely in the $z$ direction, $\beta$ is the propagation constant,
$n({\bf x}_{\rm T})$ is the (random) refractive index of the waveguide, ${\bf x}_{\rm T}$ is generally the one (two) transverse dimension(s)
in 1D (2D), $\omega=ck$, and $k=2\pi/\lambda$ where $c$ is the speed of light in vacuum. Equation~\ref{eq:helmholtz} is an eigenvalue problem in $\beta^2$ and guided modes are those
solutions (eigenfunctions) with $\beta^2>n^2_{\rm cl}k_0^2$. Here, because we consider only quasi-1D waveguides, we only have one transverse dimension, so ${\bf x}_{\rm T}=x$.
We use Dirichlet boundary condition, while noting that the choice of the boundary condition is largely inconsequential
because most guided modes strongly decay before reaching the boundary. 

In this manuscript, we do not consider the chromatic dispersion of the constituent optical materials; therefore,
all refractive indexes are assumed to be independent of the optical frequency. The reason is two-fold: first, we would like to isolate primarily the waveguiding contribution
to the dispersion, which is driven by the TAL; and second, the size of the chromatic dispersion depends on the choice of the constituent materials and only makes sense in the context of
a specific design, rather than the broad observations and arguments presented here.        
\section{Quasi-1D disordered lattice waveguide index profile}
\begin{figure}[t]
\centerline{
\includegraphics[width=\columnwidth]{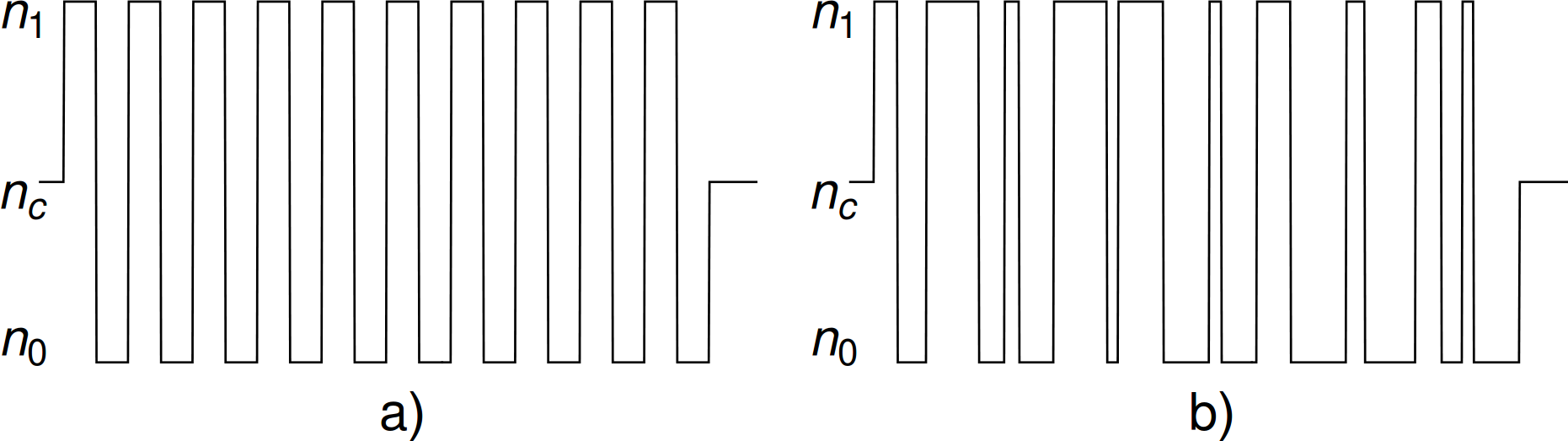}
}
\caption{\label{fig:IndexProfile}
Sample refractive index profiles of (a) ordered and (b) disordered slab waveguides are shown.}
\end{figure}

A quasi-1D ordered optical lattice waveguide can be realized by periodically stacking dielectric layers with different refractive indexes 
on top of each other. Fig.~\ref{fig:IndexProfile}(a) shows the refractive index profile of a periodic 
quasi-1D optical waveguide where $n_{0}$, $n_{1}$, and $n_{c}$ correspond to the lower index layers, higher index layers, and cladding, respectively.
We also define the refractive index contrast as $\Delta n=n_1-n_0$. In order to make a disordered waveguide, 
the thickness of the layers is randomized around an average value. We always assume that $n_{1}=1.50$ and $n_{c}=n_{0}$, while 
our simulations are carried out for either high contrast ($\Delta n=0.1$, $n_0=1.4$), or low contrast ($\Delta n=0.05$, $n_0=1.45$) .
The total number of layers alternating in $n_{0}$ and $n_{1}$ in each sample waveguide is always 100,
and the average thickness of each waveguide is $2\lambda$, where $\lambda$ is the optical wavelength at which the simulations are performed.
The average thickness is chosen for maximum localization according to Ref.~\cite{Mafi-Salman-Modal-JOSAB-2013}.
The actual thickness of each layer is a random number $2\lambda + \Delta d$, where $\Delta d$ represents the variation in the thickness.
$\Delta d$ is a random number that is chosen from a uniform random distribution of ${\rm unif}[-2r\lambda,2r\lambda]$.
The amount of randomness is controlled by the $r$-parameter, $0\le r\le 1$; where $r=0$ corresponds to the periodic lattice.
We refer to $r$ as the {\em disorder strength}. The waveguide is padded from each side with a layer of $5\lambda$ in thickness and refractive index of $n_{c}$.
Fig.~\ref{fig:IndexProfile}(b) shows a sketch of the refractive index profile of a quasi-1D disordered optical waveguide, where the number of layers are reduced 
for an easier visualization.

Note that because the total number of alternating layers is fixed at 100 but the thickness of each layer is random, the total width of
the waveguide varies in each element of the random ensemble (each waveguide). The width variation ranges from zero to 5.6\% for $0\le r\le 1$, respectively.
Although a larger value of $r$ is associated with a larger variation in the width of the waveguide, it also results in a stronger confinement,
hence making the calculations less dependent on the width of the waveguide~\cite{Mafi-Behnam-Scaling-PRB-2016}. Therefore, the very same disorder that
induces the waveguide width variation is responsible for Anderson localization, which makes the results of this paper independent of the width variation.

\begin{figure}[t]
\centerline{
\includegraphics[width=\columnwidth]{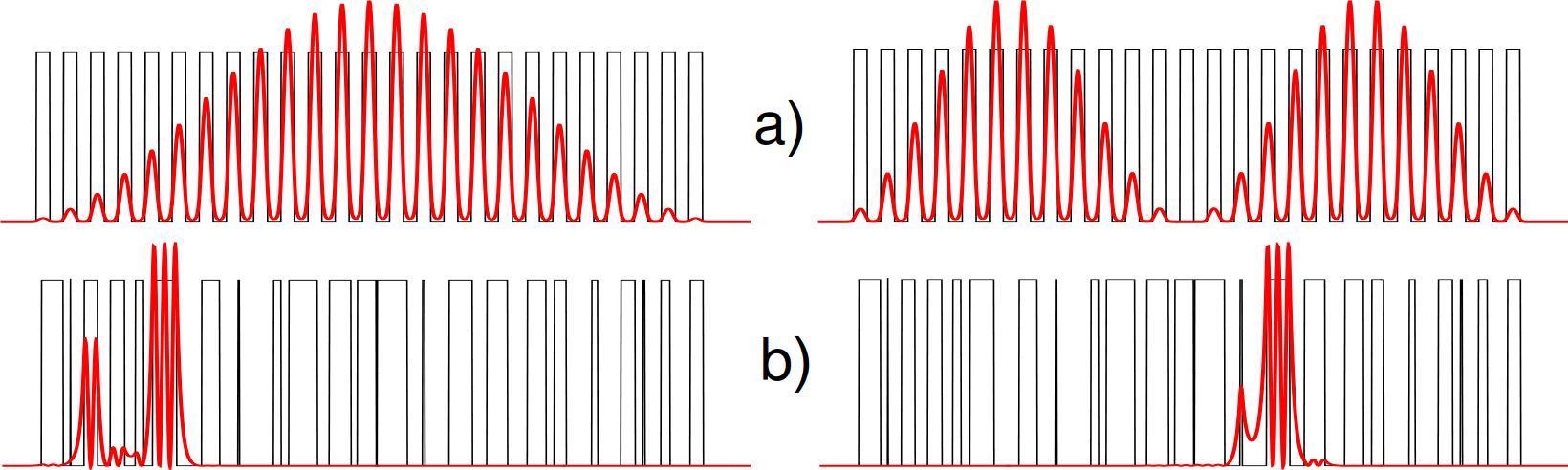}
}
\caption{\label{fig:ModeProfile}Typical mode profiles for (a) an ordered slab waveguide where each mode extends over 
the entire waveguide, and (b) a disordered slab waveguide, where the modes are spatially localized.}
\end{figure}

In Fig.~\ref{fig:ModeProfile}(a), we plot two guided modes of a quasi-1D periodic waveguide with the 
highest propagation constant. These two modes belong to a large group of standard {\em extended} Bloch periodic guided modes supported 
by the ordered optical waveguide, which are modulated by the overall mode profile of the quasi-1D waveguide~\cite{Mafi-Salman-Modal-JOSAB-2013}.
The total number of guided modes depends on the total thickness and the refractive
index values of the slabs and cladding. The key point is that each mode of the periodic structure extends over the entire 
width of the waveguide structure. A similar exercise can be done with a quasi-1D disordered waveguide, where two arbitrarily selected
modes are plotted in Fig.~\ref{fig:ModeProfile}(b) using the same refractive index parameters as that of the periodic waveguide. 
It is clear that the modes become localized in the quasi-1D disordered waveguide. While there are variations in
the shape and width of the modes, the mode profiles shown in Fig.~\ref{fig:ModeProfile}(b) are typical.

\section{Mode group index PDF}
\label{sec:PDF}
\begin{figure}[t]
\centerline{
\includegraphics[width=0.95\columnwidth]{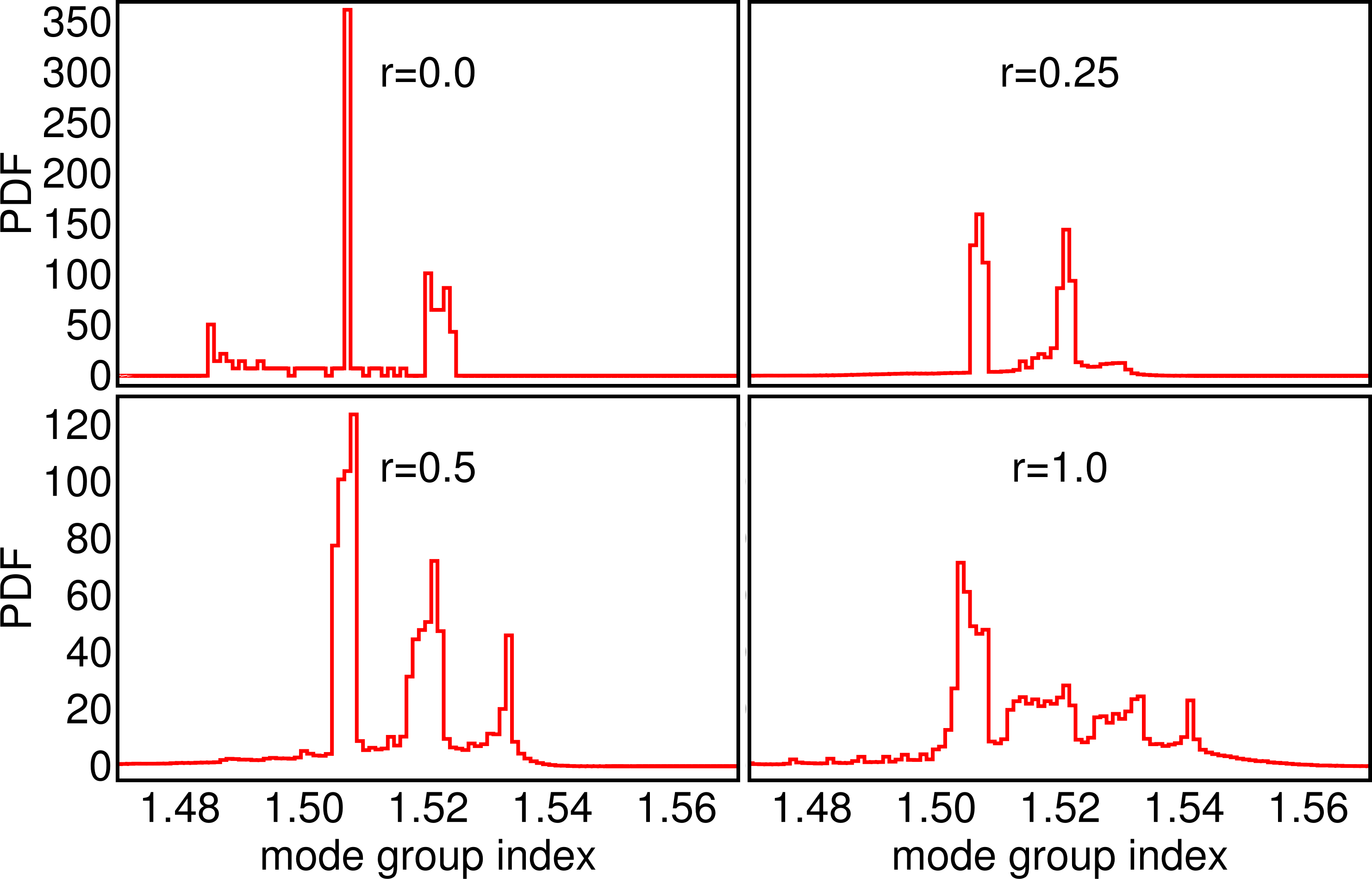}
}
\caption{\label{fig:gv-pdf-1}
Mode group index PDF for the periodic waveguide ($r=0.0$), and disordered waveguides with $r=0.25$, $r=0.50$, and $r=1.0$. Simulations are for the refractive index contrast of $\Delta n=0.1$ and the area under each 
PDF curve integrates to unity.}
\end{figure}

The group index of mode $i$ is defined as the $n^i_g=c \beta^{(1)}_i$, where $c$ is the speed of light in vacuum. In 
Fig.~\ref{fig:gv-pdf-1}, we plot the group index PDF for the periodic 
waveguide ($r=0.0$), and disordered waveguides with $r=0.25$, $r=0.50$, and $r=1.0$;
with $\Delta n=0.1$. The area under each 
PDF curve integrates to unity, and each curve is generated using the statistical information from stimulating 6,000 waveguides, amounting to
nearly 770,000 guided modes. The PDF of the periodic waveguide is highly peaked
around $n_g\approx 1.5065$; however, there are also broad secondary peaks near $n_g\approx 1.520$ and $n_g\approx 1.485$.
The modal patterns do not give any obvious clues on which category of mode shapes belong to which group-index bins.  
When a moderate amount of disorder with $r=0.25$ is introduced, the broad peak near $n_g\approx 1.485$ disappears, 
the main peak near $n_g\approx 1.520$ is lowered and broadened, while the peak around $n_g\approx 1.5065$ is raised.
As the amount of disorder is further increased to $r=0.5$ and $r=1.0$, third and fourth peaks appear at larger values
of group index, respectively. 

\begin{figure}[t]
\centerline{
\includegraphics[width=0.95\columnwidth]{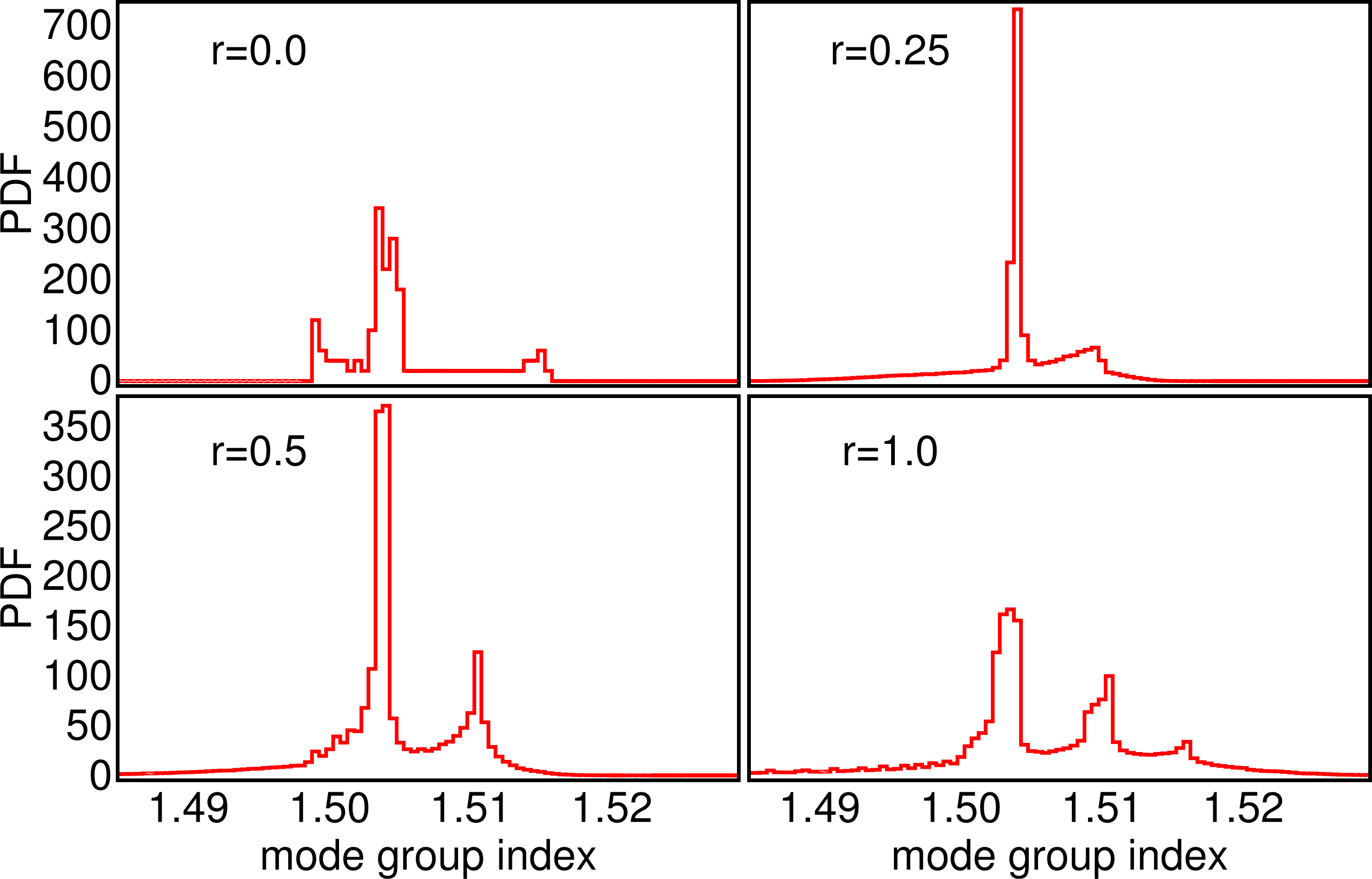}
}
\caption{\label{fig:gv-pdf-2}Similar to Fig.~\ref{fig:gv-pdf-1}, except for the refractive index contrast of $\Delta n=0.05$.}
\end{figure}

By looking at the general shapes of PDF curves in Fig.~\ref{fig:gv-pdf-1}, one can claim that a higher level of disorder amounts to a broader
PDF curve, i.e., the diversity in groups index values is increased. In other words, on average, a broader range of group velocities 
becomes accessible in the presence of increasing disorder. In order to further verify this claim, in Fig.~\ref{fig:gv-pdf-2}, we repeat the simulations of 
Fig.~\ref{fig:gv-pdf-1}, except for the lower refractive index contrast of $\Delta n=0.05$. In Fig.~\ref{fig:gv-pdf-2},
the mode group index PDF related to $r=0.25$ appears to have the narrowest and highest form. Looking at Fig.~\ref{fig:gv-pdf-2}, 
one can claim that there appears to be an optimal amount of disorder strength that narrows the range of group velocities 
and possibly reduces the pulse broadening. We will come back to these two seemingly contradictory conclusions in section~\ref{sec:broadening}, 
but for now we continue to look for other clues on the impact of disorder on the statistical behavior of group index in these waveguides.  

\begin{figure}[t]
\centerline{
\includegraphics[width=0.95\columnwidth]{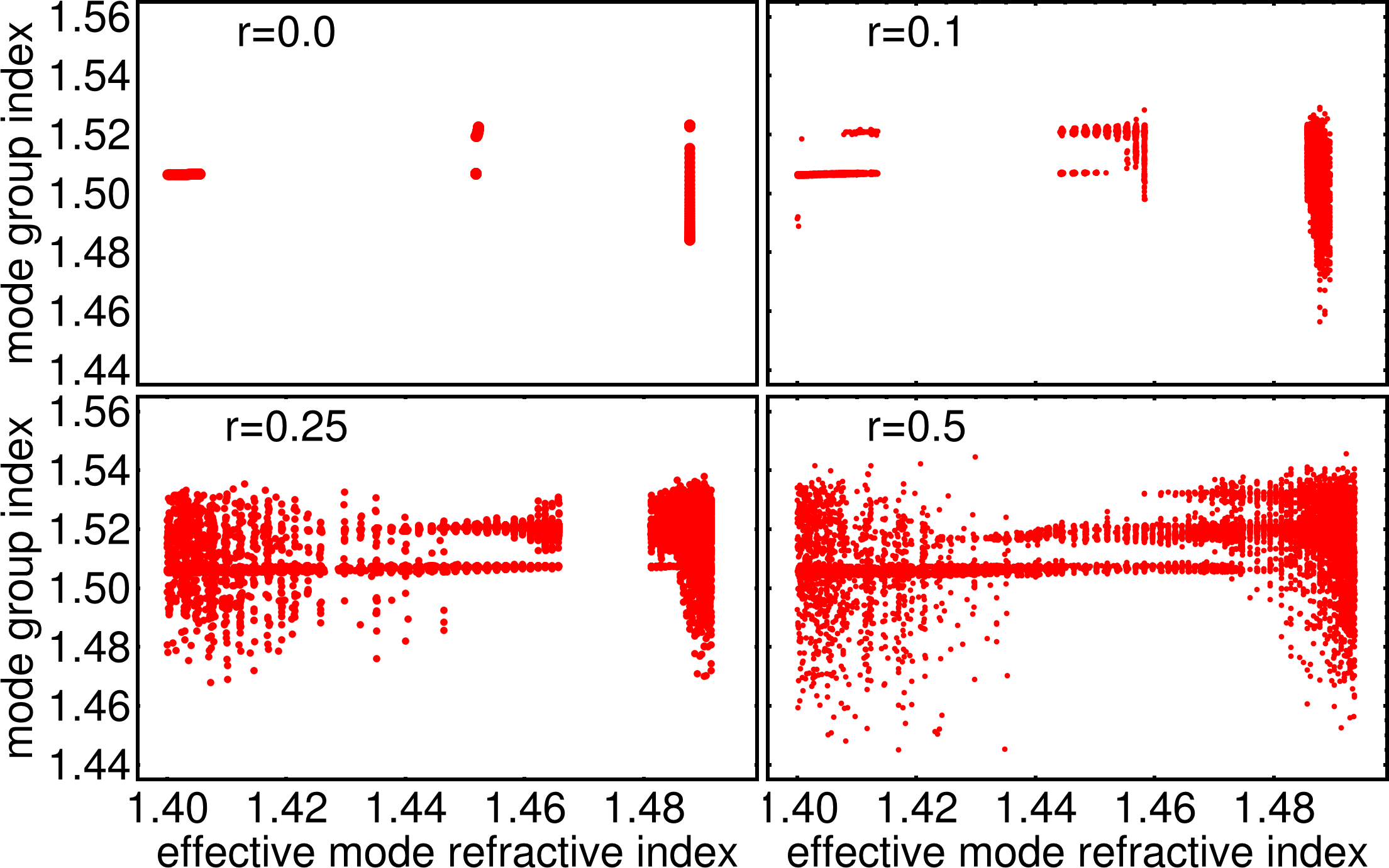}
}
\caption{\label{fig:GVD-vs-1-A} Scatter plot of the group index, $n_g$, versus the effective mode refractive index, $n_p=\beta/\omega$, for each guided mode. 
The plots are presented for $r=0.0$, $r=0.25$, $r=0.50$, and $r=1.0$. The refractive index contrast is $\Delta n=0.1$.}
\end{figure}

In Fig.~\ref{fig:GVD-vs-1-A} we present a scatter plot of the group index, $n_g$, 
versus the effective mode refractive index, $n_p=c\beta/\omega$, for each guided mode. The plots are presented again for 
$r=0.0$, $r=0.25$, $r=0.50$, and $r=1.0$, and each scatter plot is generated from stimulating 100 waveguides with the refractive index contrast of $\Delta n=0.1$. 
Note that for the periodic case of $r=0.0$, the result is always the same, because the waveguide refractive index profile is fully deterministic.
For $r=0.0$, bandgaps in $n_p$ are observed and the range of $n_g$ is limited. As the disorder is introduced and gradually increased, the
available range of both $n_p$ and $n_g$ are expanded and the gaps in $n_g$ eventually close. Again, the disorder increases the diversity in
the values of both $n_p$ and $n_g$. We recall that the accessible values of $n_p$ in a waveguide are responsible for the shape of the spatial patterns,
in addition to the {\em modal intensity profiles}. For example, if the values of $n_g$ are regularly spaced, the beam pattern in the waveguide 
repeats its shape periodically; e.g., in a graded-index optical fiber, this repetition happens with a sub-millimeter period as the beam propagates 
along the fiber~\cite{Mafi:11,liu2016kerr,wright2016self}. For disordered waveguides where such an {\rm order} is broken, pattern repetition is eliminated because
of the large number of modes and random values of $n_p$. This behavior combined with the localized {\em modal intensity profiles} is responsible 
for the high quality of image transport through TALOFs. Similar comments can be made about the diversity in values of $n_g$ and
its impact of the temporal shape of an optical pulse, which will be discussed in detail in section~\ref{sec:broadening}. 

\begin{figure}[t]
\centerline{
\includegraphics[width=0.95\columnwidth]{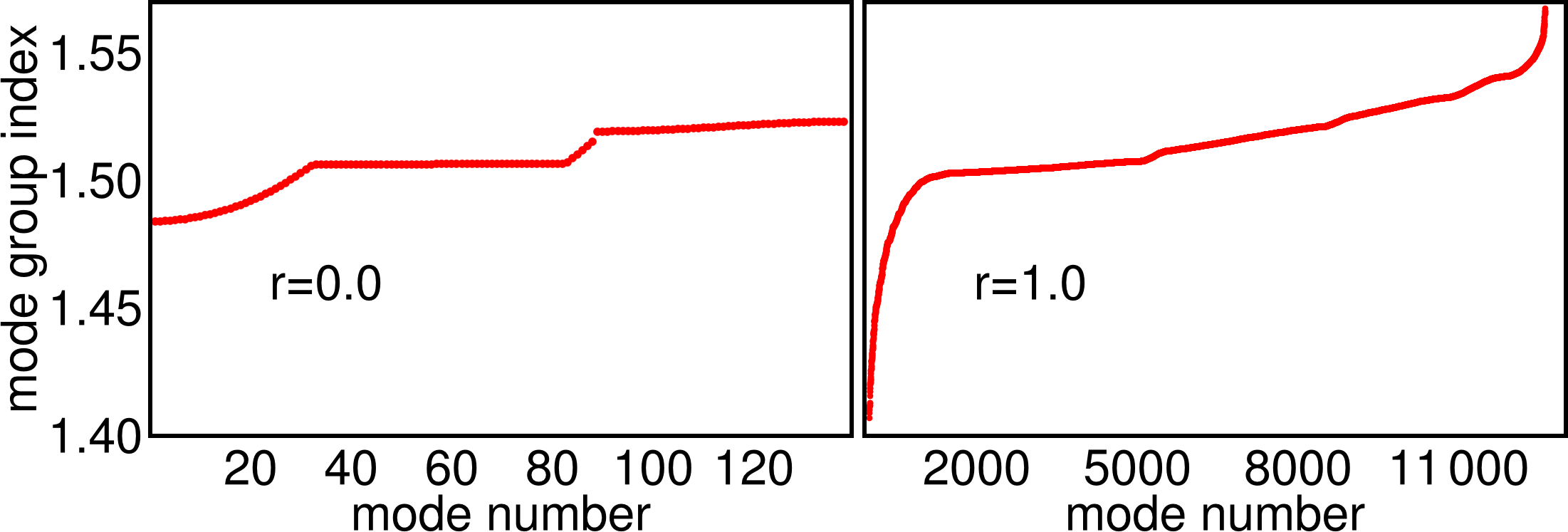}
}
\caption{\label{fig:GVD-vs-1-B}Values of the group index, $n_g$, versus the mode number for the periodic waveguide with $r=0.0$ and the maximally 
disordered waveguide $r=1.0$. For $r=1.0$, we simulate 100 waveguides and show the $n_g$ values in a ascending order. The refractive index contrast is $\Delta n=0.1$.}
\end{figure}

In order to further elaborate on the diversity of the $n_g$ values of the guided modes, in Fig.~\ref{fig:GVD-vs-1-B} we plot the
values of $n_g$ for both the periodic waveguide with $r=0.0$ and the maximally disordered waveguide $r=1.0$. The refractive index contrast is
assumed to be $\Delta n=0.1$ in this figure. In the left panel corresponding to 
$r=0.0$, we plot the values of $n_g$ versus the mode number (138 guided modes), where we have ordered the modes based on their $n_g$ values.
In the right panel corresponding to $r=1.0$, we simulate 100 waveguides and show the $n_g$ values in an ascending order. The result shows that
in a highly disordered waveguide, the group index values for the majority of the modes are still similar to those of a periodic 
waveguide with $r=0.0$; however, nearly 10\%-20\% of the modes exhibit strong deviations in group index and assume considerably smaller or larger values.
\section{Pulse propagation and broadening}
\label{sec:broadening}
The results presented in Figs.~\ref{fig:gv-pdf-1} and~\ref{fig:gv-pdf-2} show the impact of the disorder strength and refractive index contrast 
on the group index distribution in these disordered waveguides. However, it may be hard to make a concrete conclusion about pulse broadening 
from such figures, especially because the results appear to be somewhat contradictory as explained in section~\ref{sec:PDF}. 
The reason is that the pulse broadening is affected by the GV distribution 
of only those guided modes which are excited by the input pulse. For example, a typical input beam with a Gaussian spatial profile
is likely to excite those modes which have less phase variations. As we explained in section~\ref{sec:PDF}, we tried to look at the modal profiles
to build a correlation between the profiles and GV values; however, it was inconclusive. 
As such, in this section, we resort to direct computation to evaluate the impact of the GV distribution of Figs.~\ref{fig:gv-pdf-1} and~\ref{fig:gv-pdf-2} on pulse width. The pulse width
is important in setting the accessible communication bandwidth and is also essential for nonlinear properties of these waveguides.

In order to evaluate the pulse broadening, we assume that the in-coupling electric field is Gaussian in spatial profile, but is extremely 
narrow (Dirac delta function) in its temporal profile. The Gaussian spatial profile has a radius of $w=5$\,\textmu m, is 
centrally aligned with the waveguide, and is expressed as
\begin{equation}
\label{eq:EprofileSMin}
\ket{W}=
\Big(\dfrac{2}{\pi w^2}\Big)^{1/4}
\exp{\left(-\dfrac{x^2}{w^2}\right)},
\end{equation}
where $\braket{W|W}=1$. The bra-ket notation indicates integration in the transverse $x$-coordinate. The guided modes in each
waveguide are identified with $\ket{i}$ ($\braket{i|i}=1$), where $i=1,\cdots,M$ is the mode index. For example, for $\Delta n=0.1$, there
are approximately $M=140$ guided modes supported by the waveguide. For each waveguide, the modal excitation amplitudes are calculated
as $c_i=\braket{i|W}$ and the fractional power in each mode is given by $p_i=|c_i|^2$. The fractional power is nonzero mainly for 
those modes which are positioned near the center of the waveguide and have an overlap with $\ket{W}$. We define the coupling efficiency
as $\eta=\sum^M_{i=1}p_i$, where $0\leq \eta\leq 1$. When $\eta<1$, which is almost always the case, some of the power does not couple to the
guided modes and is radiated out. In order to calculate the
pulse broadening due to the modal dispersion, we follow the procedure outlined in Ref.~\cite{Pepeljugoski,Mafi-bandwidth}.
The temporal profile of the input excitation is $\delta(t)$; however, as it couples into different modes that propagate with different GVs, the pulse breaks 
into multiple subpulses:
\begin{align}
\delta(t)\to \sum^M_{i=1}p_i\delta(t-\tau_i),
\end{align}
where $\tau_i=n^i_gL/c$ is the modal delay for mode $\ket{i}$, $n^i_g$ is the group index of mode $\ket{i}$, $L$ is the propagation length, and $c$ is the
speed of light in vacuum. The pulse broadening, $\delta\tau$, is calculated using 
\begin{align}
(\delta\tau)^2=2\eta^{-1}\sum^M_{i=1}p_i(\tau_i-\bar{\tau})^2,
\end{align}
where $\bar{\tau}$ is the temporal center of the broken (broadened) pulse given by
\begin{align}
\bar{\tau}=\eta^{-1}\sum^M_{i=1}p_i\tau_i.
\end{align}

\begin{figure}[t]
\centerline{
\includegraphics[width=0.95\columnwidth]{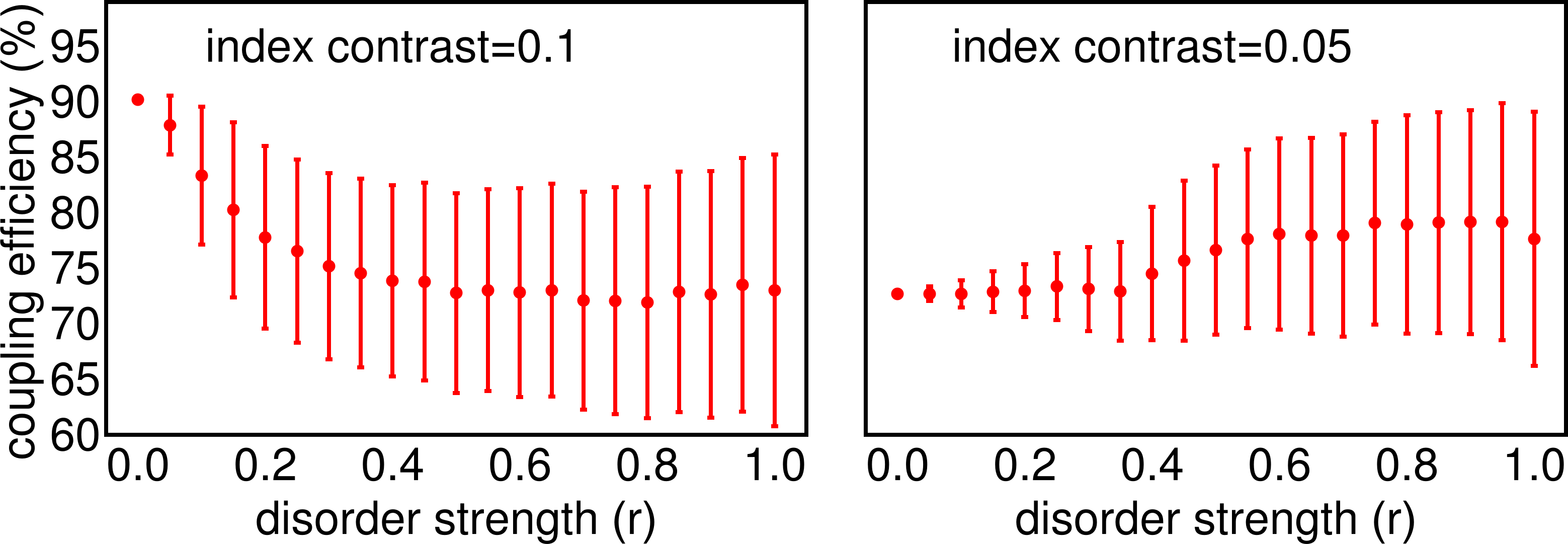}
}
\caption{\label{fig:coupling-efficiency-vs-disorder}The coupling efficiency is plotted as a function of the disorder strength for two cases
of $\Delta n=0.1$ and $\Delta n=0.05$. Each data point is averaged over 1,000 waveguides and the error-bar indicates 
one standard deviation around the mean value.}
\end{figure}

\begin{figure}[t]
\centerline{
\includegraphics[width=0.95\columnwidth]{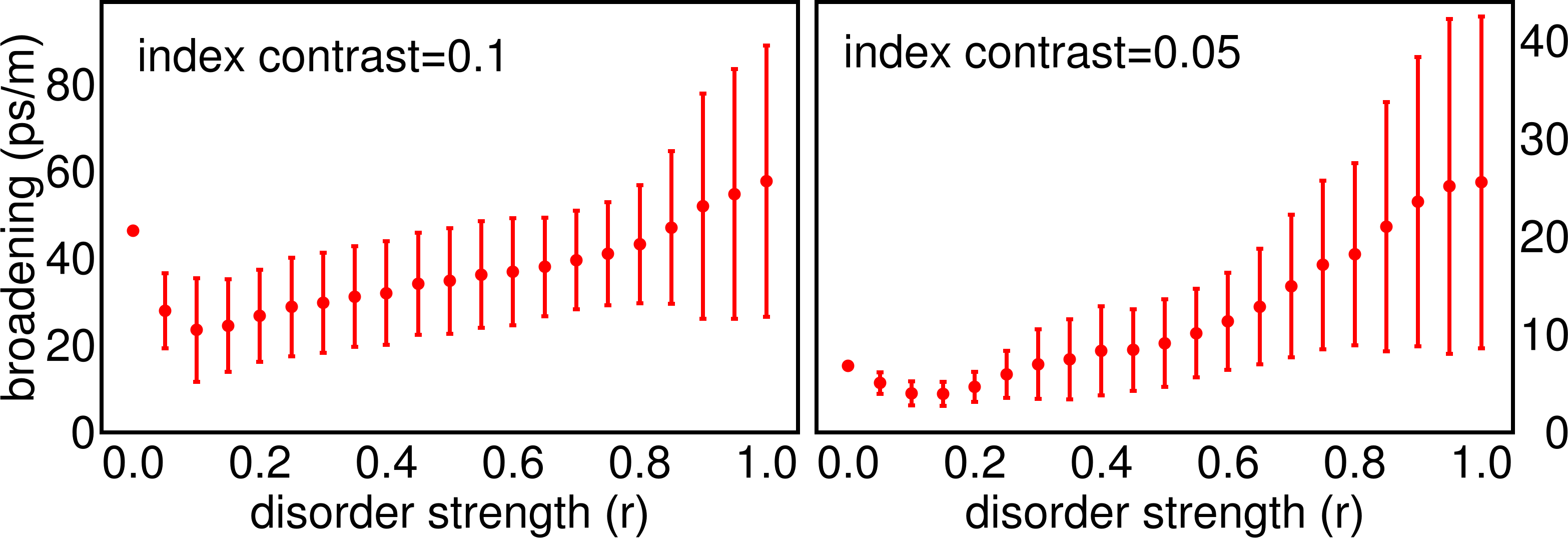}
}
\caption{\label{fig:GV-pulse-broadening-vs-disorder}Pulse broadening per unit length is plotted as a function of the disorder strength for two cases
of $\Delta n=0.1$ and $\Delta n=0.05$. Each data point is averaged over 1,000 waveguides and the error-bar indicates one standard deviation around the mean value.}
\end{figure}

In Fig.~\ref{fig:coupling-efficiency-vs-disorder}, we plot the coupling efficiency, $\eta$, as a function of the disorder strength for the two cases
of $\Delta n=0.1$ and $\Delta n=0.05$. For each data point, we have simulated 1,000 waveguides and calculated the value of $\eta$ for each 
waveguide; the dark circle shows the mean value of $\eta$ averaged over the 1,000 waveguides and the error-bar indicates 
one standard deviation around the mean value. Of course, $\eta$ is generally higher for $\Delta n=0.1$ than $\Delta n=0.05$, because a larger 
number of guided modes are supported in the former case. This result is important because it shows that in coupling to a typical disordered waveguide,
on average, only 70\%-80\% of the power can be coupled in and the rest is radiated out. In Fig.~\ref{fig:GV-pulse-broadening-vs-disorder},
we show the pulse broadening per unit length, $\delta \tau/L$, as a function of the disorder strength for two cases
of $\Delta n=0.1$ and $\Delta n=0.05$. Again, the results are averaged over the 1,000 waveguides for each data point. These figures indicate that
a small amount of disorder, typically around $r\approx 0.1-0.15$, achieves minimal pulse broadening compared to the case of no-disorder or highly
disordered waveguides. 

\begin{figure}[t]
\centerline{
\includegraphics[width=0.95\columnwidth]{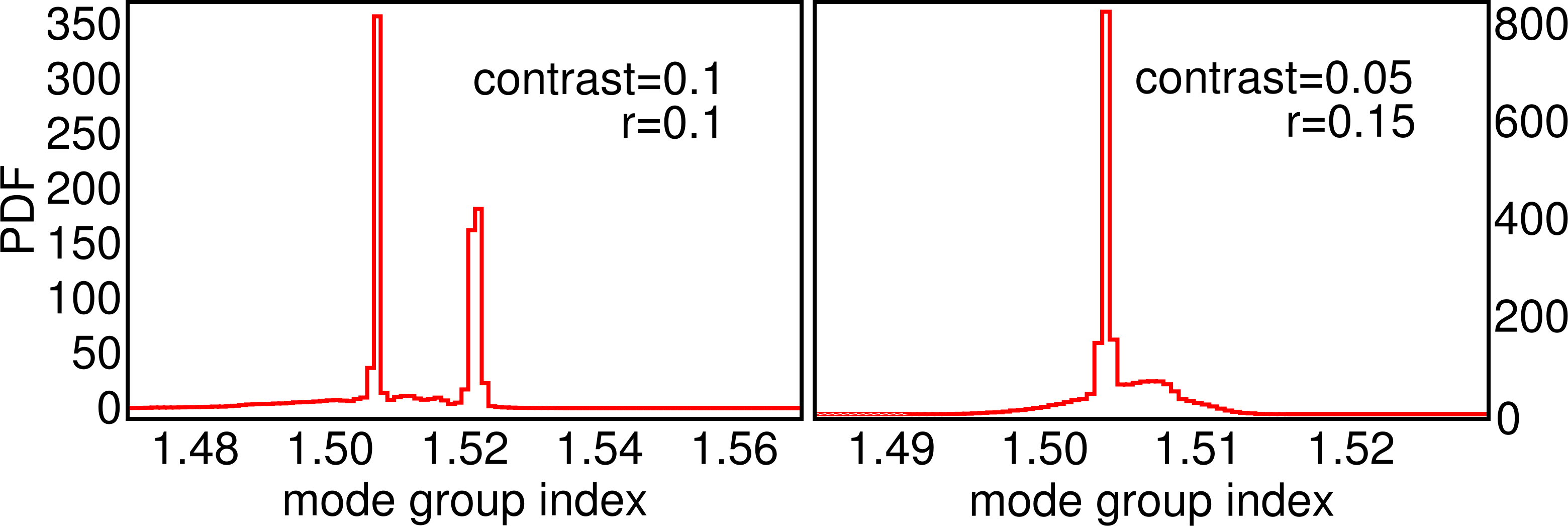}
}
\caption{\label{fig:gv-pdf-3}Similar to Fig.~\ref{fig:gv-pdf-1}, except for $\Delta n=0.1$ and $r=0.1$ in the left panel and $\Delta n=0.05$ and $r=0.15$ in the right panel.}
\end{figure}

The data from Fig.~\ref{fig:GV-pulse-broadening-vs-disorder} indicates an optimal disorder value to achieve a minimal amount of pulse broadening. Equipped with this 
information, we can now go back to the discussion surrounding Figs.~\ref{fig:gv-pdf-1} and~\ref{fig:gv-pdf-2} in section~\ref{sec:PDF}. We recall
that Fig.~\ref{fig:gv-pdf-1} indicated that a higher level of disorder amounts to a broader range of group velocities, while Fig.~\ref{fig:gv-pdf-2} indicated 
an optimal value for the disorder strength. In order to address this issue, in Fig.~\ref{fig:gv-pdf-3}, we plot the mode group index PDFs at $r=0.1$ for $\Delta n=0.1$ and 
$r=0.15$ for $\Delta n=0.05$, respectively. These values correspond to the minima of the pulse broadening curves in Fig.~\ref{fig:GV-pulse-broadening-vs-disorder}. 
By comparing Fig.~\ref{fig:gv-pdf-3} and Fig.~\ref{fig:gv-pdf-2}, it can be clearly seen that for $\Delta n=0.05$, $r=0.15$ provides the narrowest and tallest PDF curve,
which is clearly consistent with the results reported in Fig.~\ref{fig:GV-pulse-broadening-vs-disorder}. For $\Delta n=0.1$, by comparing Fig.~\ref{fig:gv-pdf-3} and Fig.~\ref{fig:gv-pdf-1},
in particular comparing $r=0.1$ and $r=0.0$, it can be observed that both the primary peak at $n_g\approx 1.5065$ and the secondary peak at $n_g\approx 1.520$ narrow down 
considerably for $r=0.1$ and the mode group index values below the primary peak disappear for $r=0.1$. As such, both results confirm the presence of an optimal value in the 
disorder strength to achieve the minimum pulse broadening.   

It must be noted that although the input pulse is assumed to be extremely narrow, i.e., a Dirac delta function, the same results would be readily obtained
with a longer input pulse. The choice of a Dirac delta function is merely a matter of convenience. In other words, for a Gaussian input pulse of temporal width $\tau_{\rm in}$,
which broadens to $\tau_{\rm out}$ upon propagation through the waveguide, it can be shown that $\delta \tau=(\tau^2_{\rm out}-\tau^2_{\rm in})^{1/2}$, where
$\delta \tau$ is independent of $\tau_{\rm in}$. Therefore, the same value of pulse broadening is obtained from a broader Gaussian pulse as from a Dirac delta function. Note that this
statement does not strictly hold if the second or higher order dispersive effects are taken into account, all of which are of higher-order-contribution and play a less important role
in the pulse broadening in normal circumstances.
\subsection{Metric for pulse broadening using the PDF curves}
\begin{figure}[t]
\centerline{
\includegraphics[width=0.95\columnwidth]{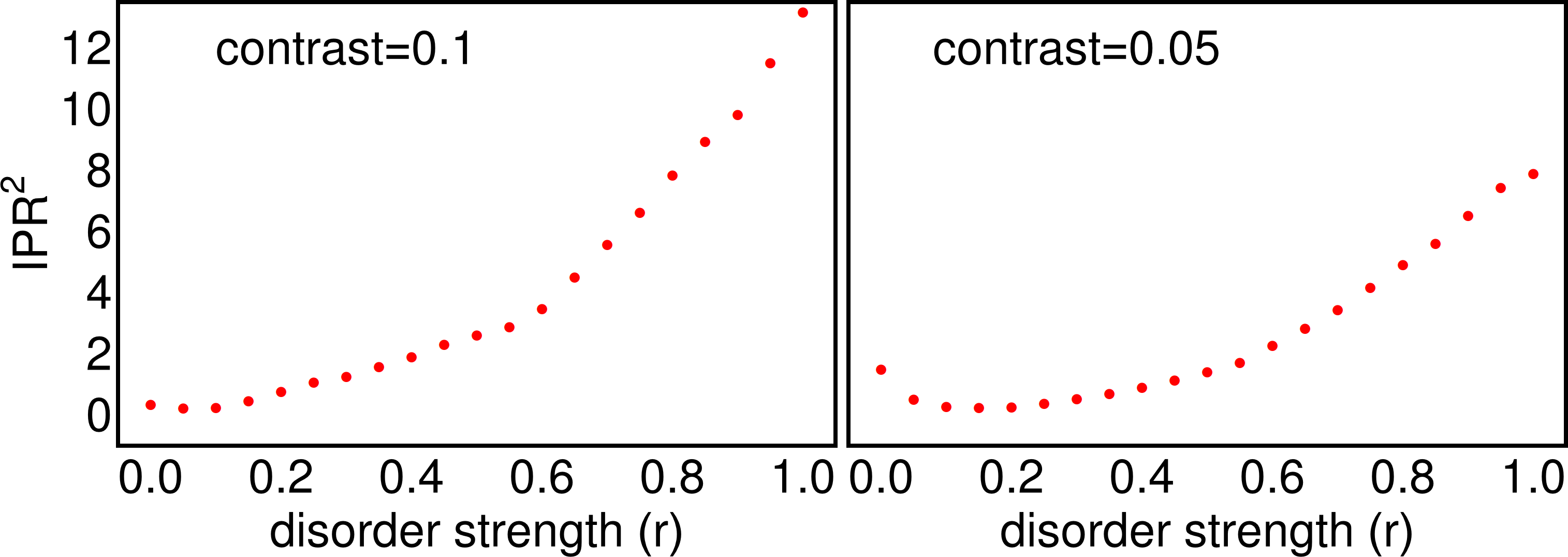}
}
\caption{\label{fig:broadening-metric}The ${\rm IPR^2}$ metric is calculated from the PDF curves as a function of the disorder contrast ($r$) for
both $\Delta n=0.1$ and $\Delta n=0.05$. Each PDF curve is obtained by simulating 1,000 waveguides.}
\end{figure}

In light of the observations in Fig.~\ref{fig:GV-pulse-broadening-vs-disorder} and how they relate to Figs.~\ref{fig:gv-pdf-1},~\ref{fig:gv-pdf-2}, and~\ref{fig:gv-pdf-3},
we define a metric to assess the width of the PDF curves; the goal is to establish a relationship between this metric and the the pulse broadening values in 
Fig.~\ref{fig:GV-pulse-broadening-vs-disorder}. A metric, at the minimum, should be able to predict the optimal of the disorder strength for minimal pulse broadening.
We use the square of the {\em inverse population ratio} (IPR) of the PDF curves as the metric. The IPR is defined as:
\begin{align}
\label{eq:defIPR}
{\rm IPR} = \int \big[\rm PDF(n_g)\big]^2\,dn_g,
\end{align}
where ${\rm PDF}(n_g)$ represents any of the PDF curves in Figs.~\ref{fig:gv-pdf-1},~\ref{fig:gv-pdf-2}, and~\ref{fig:gv-pdf-3}.
Note that unlike the commonly used 4th power in the definition of IPR (see, e.g., Ref.~\cite{PhysRevLett.105.183901}), 
we only use the 2nd power of PDF in Eq.~\ref{eq:defIPR}: the reason is that the PDF is a non-negative probability density function
and is similar to $|\psi|^2$, if $\psi$ is regarded as the (quantum-mechanical) wave amplitude. We recall that the
area under each PDF curve integrates to unity: $\int {\rm PDF}(n_g)\,dn_g=1$. In Fig.~\ref{fig:broadening-metric}, we plot the metric, i.e. ${\rm IPR^2}$
as a function of the disorder strength, both for $\Delta n=0.1$ and $\Delta n=0.05$. Fig.~\ref{fig:broadening-metric} should be compared with
Fig.~\ref{fig:GV-pulse-broadening-vs-disorder}; the disorder contrast corresponding to the minimum pulse broadening is almost exactly predicted by the metric.
Moreover, the correlation factor between the metric and the mean values presented in Fig.~\ref{fig:broadening-metric} is 89\% for $\Delta n=0.1$ and 99\% for $\Delta n=0.05$.
Therefore, the ${\rm IPR^2}$ metric appears to be a powerful tool that can predict the pulse broadening performance of such disordered waveguides directly 
using the PDF curves and without resorting to specific pulse propagation simulations.       

\section{Mode group index PDF versus localization length}
We discussed earlier that the presence of disorder results in the TAL of the guided modes. Therefore, as the disorder strength is increased, on average,
the modes should become smaller in width. In this section, we explore the correlation between the localization width of the guided modes and their group index.
For each guided mode, the mode width $\mathcal{W}$ is defined based on the standard deviation $\sigma$ of the (1D) normalized intensity 
distribution $I(x)\propto |A(x)|^2$ of the mode according to  
\begin{align}
\label{eq:sigma}
	\sigma^2 = \int_{-\infty}^{+\infty}~(x-\bar{x})^2~I(x)~dx,
\end{align}
where we define the mode center as 
\begin{align}
\label{eq:xbar}
\bar{x} &= \int_{-\infty}^{+\infty} x~I(x)~dx.
\end{align}
$x$ is the spatial coordinate across the width of the waveguide and the mode intensity profile is normalized such that $\int_{-\infty}^{+\infty} I(x) dx = 1$.
We define $\mathcal{W}=\sqrt{2}\sigma$ as a measure of the width of the modes, i.e. a larger $\mathcal{W}$ signifies a wider mode intensity profile distribution. 

\begin{figure}[t]
\centerline{
\includegraphics[width=0.95\columnwidth]{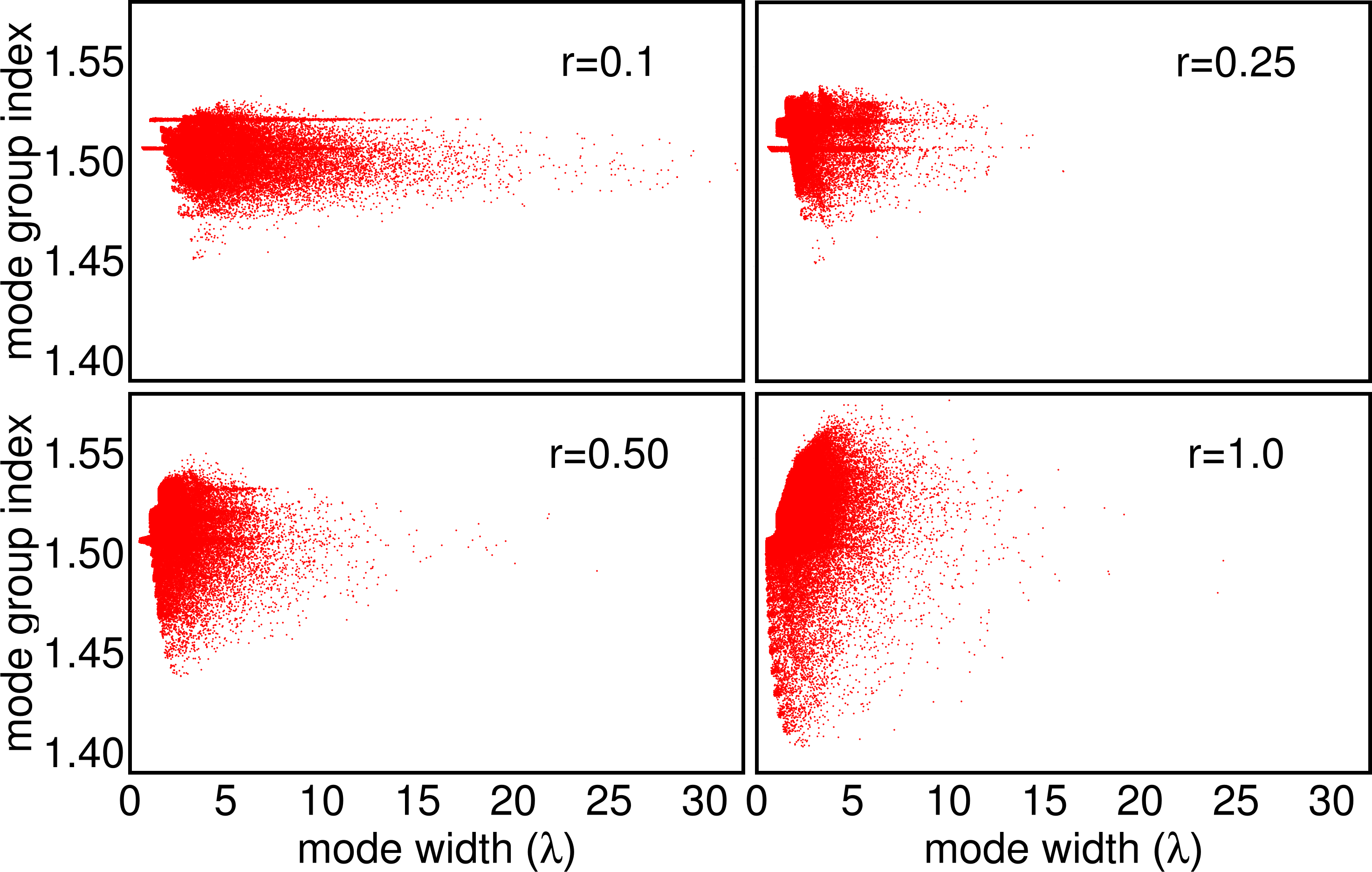}
}
\caption{\label{fig:gv-width-1}Scatter plot of the group index, $n_g$, versus the mode width, $\mathcal{W}$, for each guided mode. 
The plots are presented for $r=0.1$, $r=0.25$, $r=0.50$, and $r=1.0$. The refractive index contrast is $\Delta n=0.1$.}
\end{figure}

\begin{figure}[t]
\centerline{
\includegraphics[width=0.95\columnwidth]{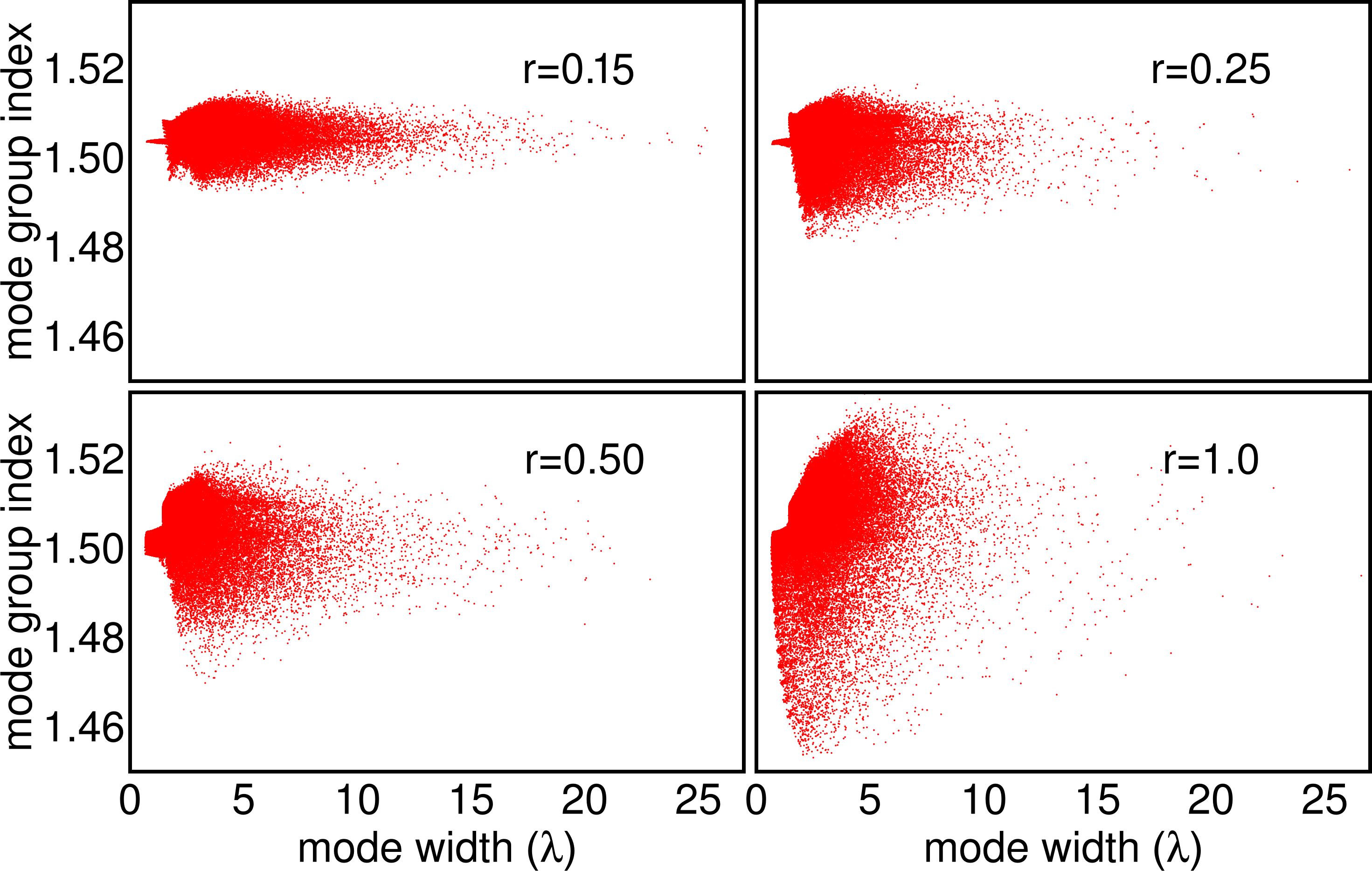}
}
\caption{\label{fig:gv-width-2}Scatter plot of the group index, $n_g$, versus the mode width, $\mathcal{W}$, for each guided mode. 
The plots are presented for $r=0.15$, $r=0.25$, $r=0.50$, and $r=1.0$. The refractive index contrast is $\Delta n=0.05$.}
\end{figure}

In Fig.~\ref{fig:gv-width-1}, we present a scatter plot of the group index, $n_g$, versus the mode width, $\mathcal{W}$, for each guided mode. 
The plots are presented for $r=0.1$, $r=0.25$, $r=0.50$, and $r=1.0$, while the refractive index contrast is assumed to be $\Delta n=0.1$. We note that
$r=0.1$ corresponds to the minimum pulse spreading for $\Delta n=0.1$ according to the left panel in Fig.~\ref{fig:GV-pulse-broadening-vs-disorder}. Similarly,
in Fig.~\ref{fig:gv-width-2}, we present a scatter plot of the group index, $n_g$, versus the mode width, $\mathcal{W}$, for each guided mode. 
The plots are presented for $r=0.15$, $r=0.25$, $r=0.50$, and $r=1.0$, while the refractive index contrast is assumed to be $\Delta n=0.05$. We note that
$r=0.15$ corresponds to minimum pulse spreading for $\Delta n=0.05$ according to the right panel in Fig.~\ref{fig:GV-pulse-broadening-vs-disorder}. The data in each
subfigure is generated from the simulation of 1,000 independent waveguides resulting in 138,000 modes. We note that the lowest value of $r$ in each figure
corresponds to the narrowest group index distribution and the widest mode-width distribution. As the disorder is increased, the group index distribution 
increases, while the mode-width distribution decreases. This observation is consistent with the TAL behavior is disordered waveguides and our discussions on group index 
distribution in previous sections.
\section{conclusion}
In this manuscript, we have introduced the mode group index PDF as a powerful tool to study the
dispersion properties of guided modes in a disordered quasi-1D slab optical waveguide.
We observe that the minimum amount of modal dispersion corresponds to a small amount of disorder, i.e., no disorder 
or large disorder both result in a large modal dispersion. We establish a metric that is applied to the 
mode group index PDF and can reliably predict the optimal amount of disorder for a minimal pulse dispersion. 
The metric is a measure of the width of the PDF and its value is strongly correlated with the modal dispersion of a pulse
propagating in the disordered waveguide. While the simulations are for a certain class of disordered quasi-1D waveguides,
they appear to conform well with the physical intuition and are likely to hold in other designs.
The results presented in the manuscript are intended to establish the framework for a comprehensive analysis of the group velocity statistics 
for quasi-2D transverse Anderson localization in disordered optical fibers in the future.  

It is quite plausible to expect that in a transversely disordered optical fiber, similar to the disordered quasi-1D slab optical waveguide,
the minimum amount of modal dispersion corresponds to an optimal (and likely a small) amount of disorder. While a longitudinally
invariant and transversely disordered optical fiber is not inherently more lossy than a conventional core-cladding optical fiber, it is likely to be
fabricated by a method that is more prone to manufacturing uncertainties, such as the stack-and-draw method~\cite{Mafi-Salman-OL-2012}.
Such manufacturing uncertainties can break the longitudinal invariance and result in attenuation, as well as polarization coupling.
The undesirable attenuation must be addressed in a case-by-case basis by making better fibers or amplifying the signal. Pulse broadening
due to the random polarization coupling is likely going to be negligible compared to the modal dispersion, similar to a conventional optical fiber;
however, this issue warrants further research.
\section*{ACKNOWLEDGMENT}
A. Mafi gratefully acknowledges support by Grant Number 1807857 from National Science Foundation (NSF).
\section*{References}
\providecommand{\noopsort}[1]{}\providecommand{\singleletter}[1]{#1}%

\end{document}